# Black Hole in Discrete Gravity

Ali H. Chamseddine[1], Ola Malaeb[1,2], and Sara Najem[1,2]

[1]Department of Physics, American University of Beirut, Beirut, Lebanon
[2]Center for Advanced Mathematical Sciences, American University of Beirut, Beirut, Lebanon

## ABSTRACT

We study the metric corresponding to a three-dimensional coset space $SO(4)/SO(3)$ in the lattice setting. With the use of three integers $n_1, n_2$, and $n_3$, and a length scale, $l_\mu$, the continuous metric is transformed into a discrete space. The numerical outcomes are compared with the continuous ones. The singularity of the black hole is explored and different domains are studied.

## 1  INTRODUCTION

The challenges encountered while attempting to quantize gravity have inspired the advancement of discrete gravity theories. Different approaches to lattice gravity were explored over the past years ([7], [8], [1]). Lately, in [5], a new approach to discrete gravity was proposed where the manifold is taken to be discrete and consists of elementary cells. The dimension $d$ is defined assuming each cell has $2d$ neighboring cells that share a common boundary with each individual cell, and a finite number of degrees of freedom is associated with each cell. This approach stands out from others primarily because it clearly reveals the continuous limit.

Recently and based on the above model, the scalar curvature of discrete gravity in two dimensions was investigated in [3], while the examination of the curvature tensor in three dimensions can be found in [4]. In the latter, a three-sphere was considered and its continuous metric was converted into a lattice. It was shown that the scalar curvature in the discrete space approaches the expected value in the continuous limit.

In this paper, we will study again the three-dimensional case, but now for a black hole coset space metric. To discretize, a length scale $l_\mu$ and three integers $n_1, n_2, n_3$ are used. The metric corresponding to a three-dimensional coset space, along with the spin connections and curvatures in the continuous case, are presented in the section 2. In section 3, we discretize the continuous metric of the black hole coset space. We investigate the domains in the proximity of the singularities and away from them, and numerically compare the discrete values of the curvature tensor with the expected continuous ones.

## 2  BLACK HOLE COSET SPACE METRIC

Consider the metric corresponding to a three-dimensional coset space $SO(4)/SO(3)$ ([6], [2])

$$ds^2 = \frac{\tanh^2 z}{u}dx^2 + \frac{\coth^2 z}{u}dy^2 + dz^2, \tag{1}$$

where

$$u = 1 - (x^2 + y^2), \quad u \geq 0, \quad -\infty \leq z \leq \infty. \tag{2}$$

A coset space of the form $G/H$ is the space of elements where $g \in G$ can be decomposed in the form:

$$g = kh \in G, \quad k = \exp\left(i\theta^{i4} J_{i4}\right) \in G/H, \quad h = \exp\left(i\theta^{ij} J_{ij}\right), \quad i,j = 1,2,3.$$

This metric was first obtained to represent a configuration of a metric in the presence of a dilaton field, but we will not add the dilaton here. Let:

$$\theta^1 = \frac{\tanh z}{\sqrt{u}} dx, \quad \theta^2 = \frac{\coth z}{\sqrt{u}} dy, \quad \theta^3 = dz$$

be the three one-forms. Then:

$$d\theta^1 = -\frac{1}{\sinh z \cosh z} \theta^1 \wedge \theta^3 - \frac{y \tanh z}{\sqrt{u}} \theta^1 \wedge \theta^2,$$

$$d\theta^2 = \frac{1}{\sinh z \cosh z} \theta^2 \wedge \theta^3 + \frac{x \coth z}{\sqrt{u}} \theta^1 \wedge \theta^2,$$

$$d\theta^3 = 0.$$

From these we get the connection one-forms:

$$\omega^{12} = \frac{y \tanh z}{\sqrt{u}} \theta^1 - \frac{x \coth z}{\sqrt{u}} \theta^2,$$

$$\omega^{13} = \frac{1}{\sinh z \cosh z} \theta^1,$$

$$\omega^{23} = -\frac{1}{\sinh z \cosh z} \theta^2$$

from which we deduce that the only non-vanishing $\omega_\mu^{ij}$ are:

$$\omega_1^{12} = \frac{y \tanh^2 z}{u}, \quad \omega_2^{12} = -\frac{x \coth^2 z}{u},$$

$$\omega_1^{13} = \frac{1}{\sqrt{u} \cosh^2 z}, \quad \omega_2^{23} = -\frac{1}{\sqrt{u} \sinh^2 z}.$$

Next, we calculate the following:

$$d\omega^{13} = \frac{2}{\cosh^2 z} \theta^1 \wedge \theta^3 - \frac{y}{\sqrt{u} \cosh^2 z} \theta^1 \wedge \theta^2$$

$$d\omega^{23} = -\frac{2}{\sinh^2 z} \theta^2 \wedge \theta^3 - \frac{x}{\sqrt{u} \sinh^2 z} \theta^1 \wedge \theta^2$$

$$d\omega^{12} = -\frac{1}{u}\left((1-x^2+y^2)\tanh^2 z + (1+x^2-y^2)\coth^2 z\right) \theta^1 \wedge \theta^2$$

$$- \frac{2y}{\sqrt{u}\cosh^2 z} \theta^1 \wedge \theta^3 - \frac{2x}{\sqrt{u}\sinh^2 z} \theta^2 \wedge \theta^3.$$

The curvature two-forms are then:

$$R^{13} = \frac{2}{\cosh^2 z}\left(\theta^1 \wedge \theta^3 - \frac{y}{\sqrt{u}} \theta^1 \wedge \theta^2\right)$$

$$R^{23} = -\frac{2}{\sinh^2 z}\left(\theta^2 \wedge \theta^3 + \frac{x}{\sqrt{u}} \theta^1 \wedge \theta^2\right)$$

$$R^{12} = \left(-2 - \frac{2}{u}\left(y^2 \tanh^2 z + x^2 \coth^2 z\right)\right) \theta^1 \wedge \theta^2 - \frac{2y}{\sqrt{u}\cosh^2 z} \theta^1 \wedge \theta^3 - \frac{2x}{\sqrt{u}\sinh^2 z} \theta^2 \wedge \theta^3.$$

Defining:

$$R^{ij} = \frac{1}{2} R_{kl}{}^{ij} \theta^k \wedge \theta^l$$

we get the non-vanishing components

$$R_{13}{}^{13} = \frac{2}{\cosh^2 z}, \quad R_{12}{}^{13} = -\frac{2y}{\sqrt{u}\cosh^2 z},$$

$$R_{23}{}^{23} = -\frac{2}{\sinh^2 z}, \quad R_{12}{}^{23} = -\frac{2x}{\sqrt{u}\sinh^2 z},$$

$$R_{13}{}^{12} = -\frac{2y}{\sqrt{u}\cosh^2 z}, \quad R_{23}{}^{12} = -\frac{2x}{\sqrt{u}\sinh^2 z}, \quad R_{12}{}^{12} = -2\left(1 + \frac{1}{u}\left(y^2 \tanh^2 z + x^2 \coth^2 z\right)\right).$$

We then define the Ricci tensor:
$$R_l^j = R_{il}{}^{ij}$$
so that:

$$R_1^1 = R_{i1}{}^{i1} = R_{21}{}^{21} + R_{31}{}^{31} = -2\left(\tanh^2 z + \frac{1}{u}\left(y^2\tanh^2 z + x^2\coth^2 z\right)\right),$$

$$R_2^1 = R_{i2}{}^{i1} = R_{32}{}^{31} = 0$$

$$R_3^1 = R_{i3}{}^{i1} = R_{23}{}^{21} = \frac{2x}{\sqrt{u}\sinh^2 z},$$

$$R_1^2 = R_{i1}{}^{i2} = R_{31}{}^{32} = 0$$

$$R_2^2 = R_{i2}{}^{i2} = R_{12}{}^{12} + R_{32}{}^{32} = -2\left(\coth^2 z + \frac{1}{u}\left(y^2\tanh^2 z + x^2\coth^2 z\right)\right),$$

$$R_3^2 = R_{i3}{}^{i2} = R_{13}{}^{12} = -\frac{2y}{\sqrt{u}\cosh^2 z},$$

$$R_1^3 = R_{i1}{}^{i3} = R_{21}{}^{23} = \frac{2x}{\sqrt{u}\sinh^2 z},$$

$$R_2^3 = R_{i2}{}^{i3} = R_{12}{}^{13} = -\frac{2y}{\sqrt{u}\cosh^2 z},$$

$$R_3^3 = R_{i3}{}^{i3} = R_{13}{}^{13} + R_{23}{}^{23} = -\frac{2}{\sinh^2 z\cosh^2 z}.$$

The scalar curvature is then

$$R = -4\left(1 + \frac{1}{\sinh^2 z\cosh^2 z} + \frac{1}{u}\left(y^2\tanh^2 z + x^2\coth^2 z\right)\right). \quad (3)$$

Since $u \geq 0$, $R$ will always have negative curvature. There is a singularity at all points on the circle defined by $u = 0$ and at $z = 0$. We can further confirm this by calculating the curvature invariant $R^{ab}R_{ab}$

$$R^{ab}R_{ab} = (R_{11})^2 + (R_{22})^2 + (R_{33})^2 + 2(R_{12})^2 + 2(R_{13})^2 + 2(R_{23})^2.$$

To relate it to the lattice setting, the metric used here gives the curvature tensor in flat coordinates (anholonomic system). This is the system where:

$$\theta^1 = e_1^1 dx, \quad \theta^2 = e_2^2 dy, \quad \theta^3 = e_3^3 dz$$

so that:

$$e_1^1 = \frac{\tanh z}{\sqrt{u}}, \quad e_2^2 = \frac{\coth z}{\sqrt{u}}, \quad e_3^3 = 1. \quad (4)$$

Thus, we will have:
$$R_{\mu\nu}{}^{ij} = e_\mu^k e_\nu^l R_{kl}{}^{ij}.$$

In terms of components, we will have:

$$R_{\underset{12}{..}}{}^{ij} = e_1^1 e_2^2 R_{12}{}^{ij}, \quad R_{\underset{13}{..}}{}^{ij} = e_1^1 e_3^3 R_{13}{}^{ij}, \quad R_{\underset{23}{..}}{}^{ij} = e_2^2 e_3^3 R_{23}{}^{ij}.$$

For the non-vanishing components, we have

$$R_{\underset{13}{..}}{}^{13} = \frac{2\sinh z}{\sqrt{u}\cosh^3 z}, \quad R_{\underset{12}{..}}{}^{13} = -\frac{2y}{u^{\frac{3}{2}}\cosh^2 z},$$

$$R_{\underset{23}{..}}{}^{23} = -\frac{2\cosh z}{\sqrt{u}\sinh^3 z}, \quad R_{\underset{12}{..}}{}^{23} = -\frac{2x}{u^{\frac{3}{2}}\sinh^2 z},$$

$$R_{\underset{13}{..}}{}^{12} = -\frac{2y\sinh z}{u\cosh^3 z}, \quad R_{\underset{23}{..}}{}^{12} = -\frac{2x\cosh z}{u\sinh^3 z}, \quad R_{\underset{12}{..}}{}^{12} = -\frac{2}{u}\left(1 + \frac{1}{u}\left(y^2\tan^2 z + x^2\cot^2 z\right)\right).$$

We note that we can compare with the discrete case by observing that:

$$R_{i\dot{3}}{}^{13} = -R_{i\dot{3}}{}^{2}, \quad R_{i\dot{2}}{}^{13} = -R_{i\dot{2}}{}^{2},$$
$$R_{2\dot{3}}{}^{23} = R_{2\dot{3}}{}^{1}, \quad R_{i\dot{2}}{}^{23} = R_{i\dot{2}}{}^{1},$$
$$R_{i\dot{3}}{}^{12} = R_{i\dot{3}}{}^{3}, \quad R_{2\dot{3}}{}^{12} = R_{2\dot{3}}{}^{3}, \quad R_{i\dot{2}}{}^{12} = R_{i\dot{2}}{}^{3}.$$

## 3 DISCRETIZITION AND NUMERICS

In order to discretize, we follow the same methodology we used in [4]. Explicitly, we define:

$$\ell_1 = \ell_2 = \ell_3 = \frac{1}{N}, \quad x = \frac{n_1}{N}, \quad y = \frac{n_2}{N}, \quad z = \frac{n_3}{N}, \tag{5}$$

where

$$n_1 = 1, 2, \cdots N, \quad n_2 = 1, 2, \cdots N, \quad n_3 = 1, 2, \cdots N, \tag{6}$$

which together with the constraint (from equation 2) give:

$$u = 1 - \frac{n_1^2 + n_2^2}{N^2} > 0. \tag{7}$$

To derive the three-dimensional discrete curvature, we start with the definition [5]

$$\frac{i}{2} R^i_{\mu\nu} \sigma^i = \frac{1}{2\ell^\mu \ell^\nu} \left( \Omega_\mu(n) \Omega_\nu(n+\widehat{\mu}) \Omega_\mu^{-1}(n+\widehat{\nu}) \Omega_\nu^{-1}(n) - \mu \leftrightarrow \nu \right)$$

$$= \frac{1}{2\ell^\mu \ell^\nu} \left\{ \left( \cos \frac{1}{2} \ell^\mu \omega_\mu(n) + i\widehat{\omega}^i_\mu(n) \sin \frac{1}{2} \ell^\mu \omega_\mu(n) \sigma^i \right) \right.$$

$$\cdot \left( \cos \frac{1}{2} \ell^\nu \omega_\nu(n+\widehat{\mu}) + i\widehat{\omega}^i_\nu(n) \sin \frac{1}{2} \ell^\nu \omega_\nu(n+\widehat{\mu}) \sigma^i \right)$$

$$\cdot \left( \cos \frac{1}{2} \ell^\mu \omega_\mu(n+\widehat{\nu}) - i\widehat{\omega}^i_\mu(n+\widehat{\nu}) \sin \frac{1}{2} \ell^\mu \omega_\mu(n+\widehat{\nu}) \sigma^i \right)$$

$$\left. \cdot \left( \cos \frac{1}{2} \ell^\nu \omega_\nu(n) - i\widehat{\omega}^i_\nu(n) \sin \frac{1}{2} \ell^\nu \omega_\nu(n) \sigma^i \right) - \mu \leftrightarrow \nu \right\}.$$

Consider the product:

$$\left( \cos \frac{1}{2} \ell^\mu \omega_\mu(n) + i\widehat{\omega}^i_\mu(n) \sin \frac{1}{2} \ell^\mu \omega_\mu(n) \sigma^i \right) \left( \cos \frac{1}{2} \ell^\nu \omega_\nu(n+\widehat{\mu}) + i\widehat{\omega}^i_\nu(n+\widehat{\mu}) \sin \frac{1}{2} \ell^\nu \omega_\nu(n+\widehat{\mu}) \sigma^i \right),$$

this can be rewritten as:

$$\left( \cos \frac{1}{2} \ell^\mu \omega_\mu(n) \cos \frac{1}{2} \ell^\nu \omega_\nu(n+\widehat{\mu}) - \widehat{\omega}^i_\mu(n) \widehat{\omega}^i_\nu(n+\widehat{\mu}) \sin \frac{1}{2} \ell^\mu \omega_\mu(n) \sin \frac{1}{2} \ell^\nu \omega_\nu(n+\widehat{\mu}) \right)$$

$$+ i \left( \cos \frac{1}{2} \ell^\mu \omega_\mu(n) \widehat{\omega}^i_\nu(n+\widehat{\mu}) \sin \frac{1}{2} \ell^\nu \omega_\nu(n+\widehat{\mu}) + \widehat{\omega}^i_\mu(n) \sin \frac{1}{2} \ell^\mu \omega_\mu(n) \cos \frac{1}{2} \ell^\nu \omega_\nu(n+\widehat{\mu}) \right.$$

$$\left. \varepsilon^{ijk} \widehat{\omega}^j_\mu(n) \sin \frac{1}{2} \ell^\mu \omega_\mu(n) \widehat{\omega}^k_\nu(n+\widehat{\mu}) \sin \frac{1}{2} \ell^\nu \omega_\nu(n+\widehat{\mu}) \right) \sigma^i$$

$$\equiv A_{\nu\mu} + i B^i_{\mu\nu} \sigma^i,$$

where

$$A_{\mu\nu}(n) = \left( \cos \frac{1}{2} \ell^\mu \omega_\mu(n+\widehat{\nu}) \cos \frac{1}{2} \ell^\nu \omega_\nu(n) - \widehat{\omega}^j_\mu(n+\widehat{\nu}) \widehat{\omega}^j_\nu(n) \sin \frac{1}{2} \ell^\mu \omega_\mu(n+\widehat{\nu}) \sin \frac{1}{2} \ell^\nu \omega_\nu(n) \right),$$

and:

$$B^i_{\mu\nu}(n) = \left( \widehat{\omega}^i_\mu(n) \sin \frac{1}{2} \ell^\mu \omega_\mu(n) \cos \frac{1}{2} \ell^\nu \omega_\nu(n+\widehat{\mu}) + \widehat{\omega}^i_\nu(n+\widehat{\mu}) \sin \frac{1}{2} \ell^\nu \omega_\nu(n+\widehat{\mu}) \cos \frac{1}{2} \ell^\mu \omega_\mu(n) \right.$$

$$\left. -\varepsilon^{ijk} \widehat{\omega}^j_\mu(n) \sin \frac{1}{2} \ell^\mu \omega_\mu(n) \widehat{\omega}^k_\nu(n+\widehat{\mu}) \sin \frac{1}{2} \ell^\nu \omega_\nu(n+\widehat{\mu}) \right).$$

Similarly, the next pair gives:

$$\left(\cos\frac{1}{2}\ell^\mu\omega_\mu(n+\hat{v}) - i\widehat{\omega}_\mu^i(n+\hat{v})\sin\frac{1}{2}\ell^\mu\omega_\mu(n+\hat{v})\sigma^i\right)\left(\cos\frac{1}{2}\ell^v\omega_v(n) - i\widehat{\omega}_v^i(n)\sin\frac{1}{2}\ell^v\omega_v(n)\sigma^i\right)$$
$$= A_{\mu v}(n) - iB_{v\mu}^i(n).$$

Thus, the total product is:

$$\frac{i}{2}R_{\mu v}^i\sigma^i = \left(\frac{1}{2\ell^\mu\ell^v}\left(A_{v\mu} + iB_{\mu v}^i\sigma^i\right)\left(A_{\mu v} - iB_{v\mu}^i\sigma^i\right) - \mu\leftrightarrow v\right)$$
$$= \frac{i}{\ell^\mu\ell^v}\left(A_{\mu v}B_{\mu v}^i - A_{v\mu}B_{v\mu}^i + i\varepsilon^{ijk}B_{\mu v}^j B_{v\mu}^k\right)\sigma^i,$$

giving the below result:

$$R_{\mu v}^i(n) = \frac{2}{\ell^\mu\ell^v}\left(A_{\mu v}(n)B_{\mu v}^i(n) - A_{v\mu}(n)B_{v\mu}^i(n) + \varepsilon^{ijk}B_{\mu v}^j(n)B_{v\mu}^k(n)\right). \tag{8}$$

The connection $\omega_v(n)$ is determined from the zero torsion condition:

$$T_{\mu v}(n) = \frac{1}{\ell^\mu}\left(\Omega_\mu(n)e_v(n+\hat{\mu})\Omega_\mu^{-1}(n) - e_v(n)\right) - \mu\leftrightarrow v,$$

and this can be computed exactly:

$$0 = \frac{1}{\ell^\mu}\left(\left(\cos\frac{1}{2}\ell^\mu\omega_\mu(n) + i\widehat{\omega}_\mu^i(n)\sin\frac{1}{2}\ell^\mu\omega_\mu(n)\sigma^i\right)\sigma^k\left(\cos\frac{1}{2}\ell^\mu\omega_\mu(n) - i\widehat{\omega}_\mu^j(n)\sin\frac{1}{2}\ell^\mu\omega_\mu(n)\sigma^j\right)e_v^k(n+\hat{\mu})\right.$$
$$\left. - e_v^i(n)\sigma^i\right) - \mu\leftrightarrow v\right).$$

By expanding and grouping terms we get the following result:

$$T_{\mu v}^{\ i}(n) = \frac{1}{\ell^\mu}\left(\cos\ell^\mu\omega_\mu(n)e_v^i(n+\hat{\mu}) - \varepsilon^{ijk}\sin\ell^\mu\omega_\mu(n)\widehat{\omega}_\mu^j(n)e_v^k(n+\hat{\mu})\right.$$
$$\left. + 2\widehat{\omega}_\mu^i(n)\widehat{\omega}_\mu^j(n)\sin^2\frac{1}{2}\ell^\mu\omega_\mu(n)e_v^j(n+\hat{\mu}) - e_v^i(n)\right) - (\mu\leftrightarrow v).$$

The vanishing of $T_{\mu v}^{\ i}$ provides 9 conditions to solve for the 9 unknowns $\omega_\mu^i(n)$. The values of the spin connections are obtained numerically, and hence, the three-dimensional discrete curvatures are obtained using equation (8).

Equation 3 gives the expression of the scalar curvature in the continuous case. We are going to compare it with what we get from the discrete case. In the latter, the expression of the scalar curvature is given by:

$$R = 2\left(\frac{1}{e_1^1 e_2^2}R_{12}^3 + \frac{1}{e_2^2 e_3^3}R_{23}^1 + \frac{1}{e_3^3 e_1^1}R_{31}^2\right),$$

which holds when $e_\mu^i$ is diagonal, which is the case here as it is evident from equation 4. It is clear from 3 that there is a singularity at $z = 0$ and at all points on the circle defined by $u = 0$. First, we will examine the domain where $z$ is large, followed by an analysis of the domain where $z$ is small.

### 3.1 The limit of large $z$

$$R \to -\frac{4}{u}\left(1 + \frac{x^2 + y^2}{u}\right). \tag{9}$$

The singularity at $u = 0$ is evident. Therefore, in order to determine the permissible range over which the curvature is well-behaved, we define $1 - x^2 - y^2 > \varepsilon$, the range of coordinates in the vicinity of the singularity. We note that $\varepsilon = 0$ is the locus of $u = 0$ or equivalently $x^2 + y^2 = 1$, and thus finding the right value of $\varepsilon$ is equivalent to finding the proper cut-off to avoid the singularity. Therefore, we iterate over different values of $\varepsilon$ between 0 and 0.5 in increments of 0.05, and calculate the mean-square error between the continuous value of the scalar curvature and its discrete numerical counterpart defined to be $\frac{1}{n}\sum_{i}^{n}(R_{i,con} - R_{i,numerical})^2$. We repeat the process for different values of $N$. The results are plotted in figure 1. The different colors correspond to different values of $N$. We conclude from the plot that the error is converging to zero for $\varepsilon = 0.15$.

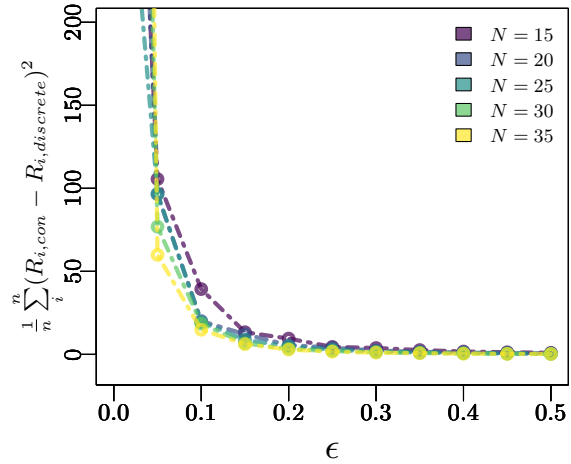

**Figure 1.** The rms error between the continuous and discrete values of the scalar curvature in the limit of large $z$ are plotted as a function of $\varepsilon$ for different $N$, where $\varepsilon$ denotes the distance away from $u = 0$ where the discretization is expected to hold.

For $\varepsilon$ smaller than 0.15, we are close to the singularity ($u = 0$) and the discretization method fails to agree with the continuous limit as the RMS error between $R_{con}$ and $R_{discrete}$ blows up. In the limit where they are in agreement, we show a sample in figure 2 for $N = 35$ and $\varepsilon = 0.15$, where we plot the values of $R_{con}$ and those of $R_{discrete}$ on the same graph with the index $i$ denoting the element number in the curvature vector associated with given coordinates $x_i, y_i$ and $z_i$.

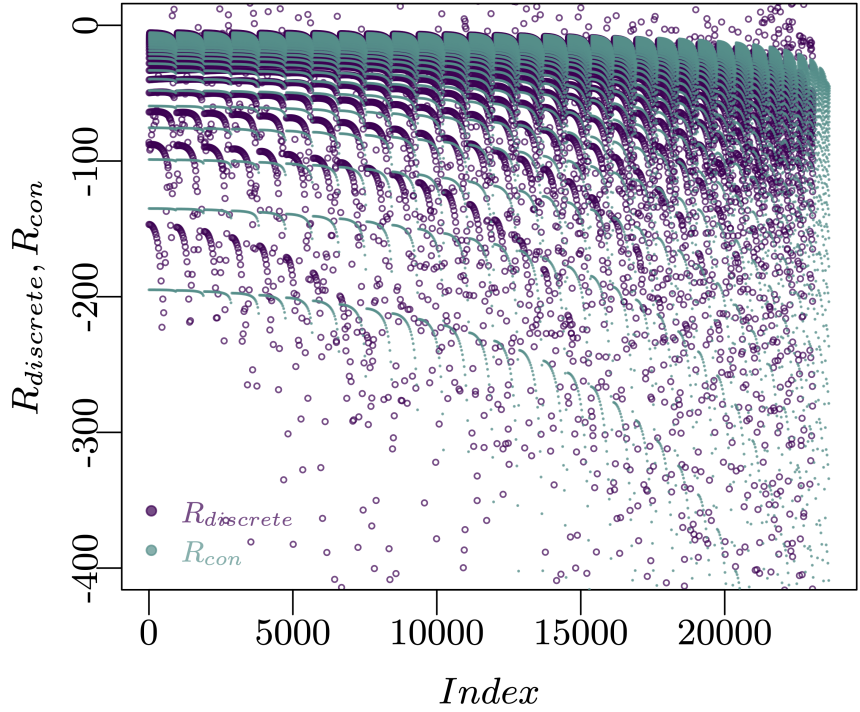

**Figure 2.** *R* continuous and discrete for $N = 35$ and $\varepsilon = 0.15$.

### 3.2 The limit of small $z$

For a small value of $z$, we are in the region near the singularity $z = 0$. Therefore, and following the same reasoning we did to find $\varepsilon$ near $u = 0$, we limit $z$ between $\varepsilon_z/N$ and 1 and follow the RMS between the continuous and discrete values of the scalar curvature. In Figure 3, we show the RMS for $N = 35$ and $\varepsilon = 0.15$, which allows us to define the lower cutoff near $z = 0$ below which the scalar curvature will blow up.

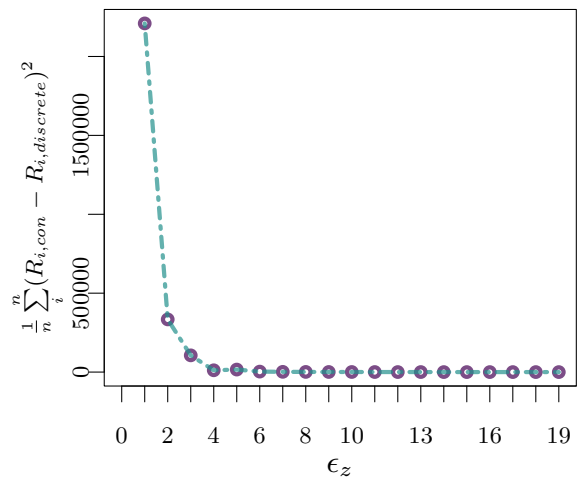

**Figure 3.** The RMS error is followed as a function of $\varepsilon_z$ for $N = 35$ and $\varepsilon = 0.15$.

It is evident from the plot that as $\varepsilon_z$ approaches 5, the error tends to zero. Further, in order to show the well-behavedness of the discrete scalar curvature, we follow it together with its continuous counterpart in this domain. Figure 4 shows the finiteness of $R_{discrete}$ as opposed to the large, near-singular behavior of $R_{con}$. The inset shows their agreement away from the jumps.

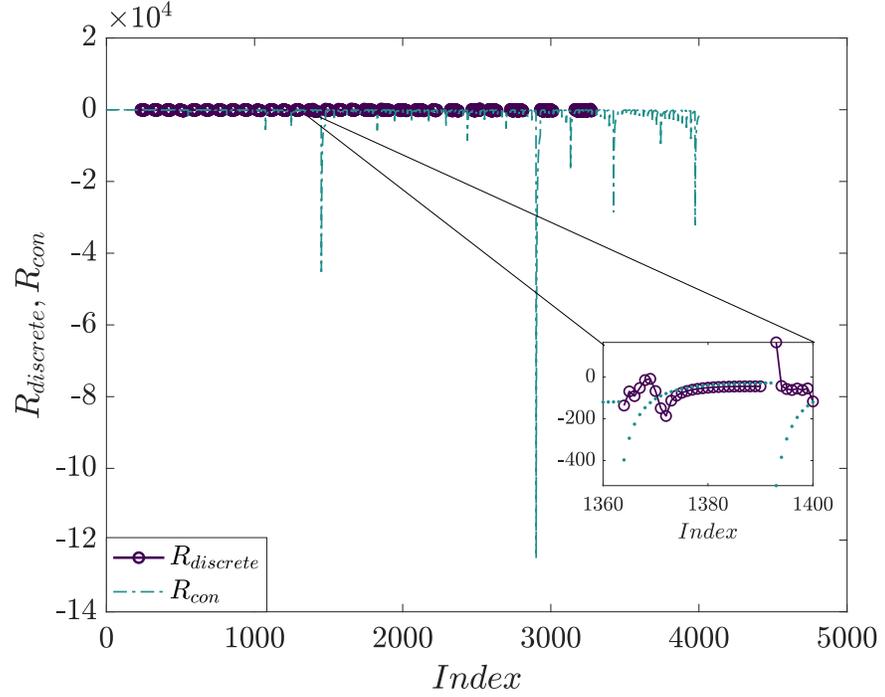

**Figure 4.** $R$ continuous and discrete are followed for $N = 35$, $\varepsilon = 0.15$, $z < 1$, and $\varepsilon_z = 5$. The inset shows their local agreement and makes it obvious that $R_{con}$ exhibits large near-singular jumps, which are avoided in the discrete case.

For very small $z$, we have $\sinh z \to z$ and $\cosh z \to 1$,

$$R \to -4\left(1 + \frac{1}{z^2} + \frac{1}{u}\left(y^2 z^2 + \frac{x^2}{z^2}\right)\right) \tag{10}$$

$$\simeq -4\left(\frac{1-y^2}{uz^2}\right) \tag{11}$$

with obvious asymmetry. Figure 5 is obtained by plotting $R_{discrete}$ and $R_{limit} = -4\left(\frac{1-y^2}{uz^2}\right)$ for small values of $z$, showing their agreement.

Further, when $x = 0$ we get $R \to -4N^2$, displaying clearly the singularity of the black hole. In Figure 6, the discrete value of the scalar curvature is plotted for $z \simeq \frac{1}{N}$ and $x = 0$ along with its expected limiting behavior, which is $-4N^2$.

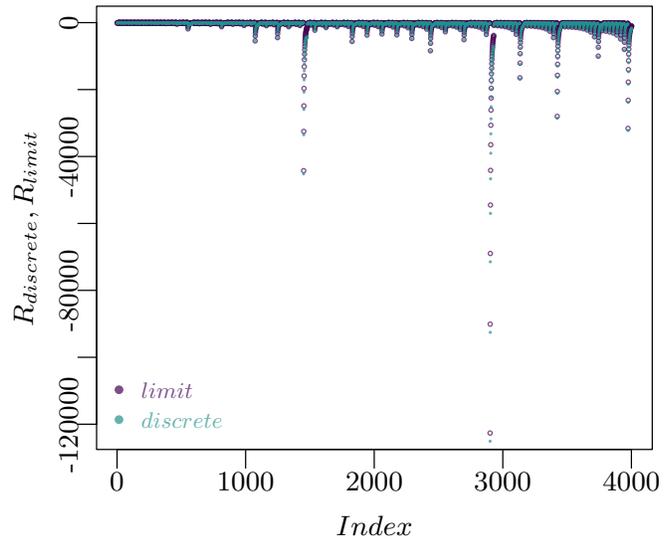

**Figure 5.** $R_{discrete}$ is compared with its limiting behavior $4(\frac{1-y^2}{uz^2})$ for small $z$.

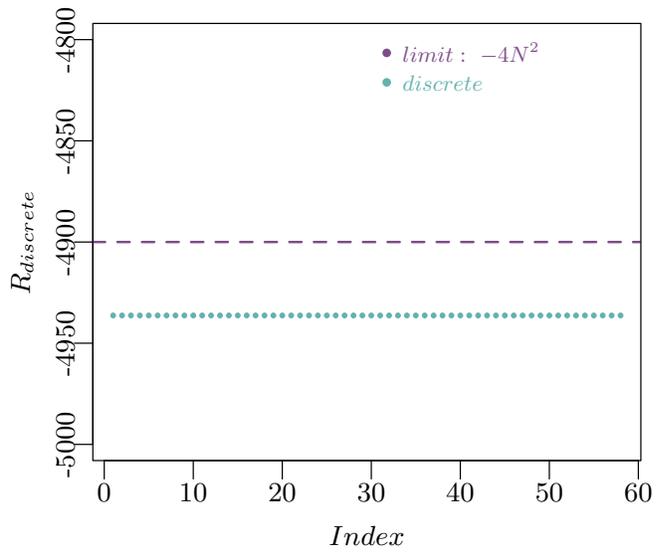

**Figure 6.** $R_{discrete}$ is compared with its limiting value $-4N^2$ for $z \approx \frac{1}{N}$ and $x = 0$.

9/10

# 4 CONCLUSION

In this paper, we applied the methodology used in [4] to define the curvature of the discrete space in the proposed model of discrete gravity. We considered the black hole coset space metric. The singularities were studied, and the curvatures in the discrete and continuous limits were compared close and far away from the singularity. We found that as we move away from the singularity, the rms error between the two curvatures tends to zero. We also showed that near the singularity, the discrete method is more reliable than the continuous.

# ACKNOWLEDGMENTS

The work of A. H. C. is supported in part by the National Science Foundation Grant No. Phys-2207663.

# SUPPORTING MATERIAL

The script used in the paper is available at the following link.